\documentclass[twocolumn,showpacs,twoside,10pt,aps,superscriptaddress]{revtex4-1}
\usepackage{makecell}
\usepackage{float}
\usepackage{graphicx}
\usepackage{dcolumn}
\usepackage{bm}
\usepackage[colorlinks=true,linkcolor=blue,citecolor=blue,urlcolor=blue]{hyperref}%
\begin{document}

\preprint{APS/123-QED}

\title{Online optimization for optical readout of a single electron spin in diamond}
\author{Xue Lin}
\affiliation{School of Microelectronics and School of Physics, Hefei University of Technology, Hefei, Anhui 230009, China}%
\affiliation{Research Center for Quantum Sensing, Zhejiang Laboratory, Hangzhou, 311000, China}%

\author{Jingwei Fan}%
\affiliation{School of Microelectronics and School of Physics, Hefei University of Technology, Hefei, Anhui 230009, China}%

\author{Runchuan Ye}%
\affiliation{School of Microelectronics and School of Physics, Hefei University of Technology, Hefei, Anhui 230009, China}%
\affiliation{Research Center for Quantum Sensing, Zhejiang Laboratory, Hangzhou, 311000, China}%

\author{Mingti Zhou}%
\affiliation{Research Center for Quantum Sensing, Zhejiang Laboratory, Hangzhou, 311000, China}%

\author{Yumeng Song}%
\affiliation{School of Microelectronics and School of Physics, Hefei University of Technology, Hefei, Anhui 230009, China}%
\affiliation{Research Center for Quantum Sensing, Zhejiang Laboratory, Hangzhou, 311000, China}%

\author{Dawei Lu}%
\email{ludw@sustech.edu.cn}
\affiliation{Shenzhen Institute for Quantum Science and Engineering and Department of Physics, Southern University of Science and Technology, Shenzhen 518055, China}%

\author{Nanyang Xu}%
\email{nyxu@zhejianglab.edu.cn}
\affiliation{Research Center for Quantum Sensing, Zhejiang Laboratory, Hangzhou, 311000, China}%
\date{\today}

\begin{abstract}
The nitrogen-vacancy (NV) center in diamond has been developed as a promising platform for quantum sensing, especially for magnetic field measurements in the nano-tesla range with a nano-meter resolution. Optical spin readout performance has a direct effect on the signal-to-noise ratio (SNR) of experiments. In this work, we introduce an online optimization method to customize the laser waveform for readout. Both simulations and experiments reveal that our new scheme optimizes the optically detected magnetic resonance in NV center. The SNR of optical spin readout has been witnessed a 44.1\% increase in experiments. In addition, we applied the scheme to the Rabi oscillation experiment, which shows an improvement of 46.0\% in contrast and a reduction of 12.1\% in mean deviation compared to traditional constant laser power SNR optimization. This scheme is promising to improve sensitivities for a wide range of NV-based applications in the future.
\end{abstract}

\maketitle


\section{Introduction}
The sensing of weak signals with high spatial resolution is of great importance in diverse areas ranging from fundamental physics and material science to biological sciences. The nitrogen-vacancy (NV) center in diamond is one of the major platforms in the emerging field of quantum technologies because of its remarkable stability, long coherence time, and excellent controllability via microwave or optical methods. The level structure and coherence time are sensitive to different physical fields, which makes it a multi-functional sensor with nano-scale spatial resolution and high sensitivity. Up to now, it has been applied in sensing magnetic fields \cite{Wrachtrup2008NanoscaleImagingMagnetometry,Wrachtrup2017BroadbanMmagneticFields,Lukin2008NanoscaleMagneticSensing,Wrachtrup2016Sensitivity,Fedor2016Metrolog,Degen2008MagneticField,Jing-WeiZhou2014metrology}, electric fields \cite{Wrachtrup2011ElectricField}, temperature \cite{Lukin2013Thermometry,D.2013Thermometry}, stress \cite{stress2019} and magnetic resonance imaging \cite{Wrachtrup2013MagneticResonanceSpectroscopy,ShiFazhan2015ResonanceSpectroscopy,Jacques2014NanoscaleImaging,Degen2009NanoscaleImaging}.

For NV-based sensing applications, a common control sequence consisting of spin initialization, manipulation, and readout is typically followed. Due to the control imperfections, the performance of current NV-based quantum sensors is hard to reach their theoretical limitation \cite{Degen2017quantumsensing,barry2020sensitivity}. To achieve the potential sensitivities in different applications \cite{barry2016magnetic,davis2018magnetic}, further improving the performance of NV-based sensors is a flourishing and multi-disciplinary research topic \cite{hopper2016ChargeControlAndSpinReadout,neumann2010SingleShot,qian2021machine}. Among these problems, the optical pumping process that refers to the spin initialization and readout is of particular importance to improve the signal-to-noise ratio (SNR) of sensing applications. Many efforts have been made to optimize the optical pumping process via changing the laser shape or power. For example, Song $et$ $al$. found that extremely short laser pulses are beneficial for electron spin polarization when it is repetitively applied \cite{song2020pulse}. Liu $et$ $al$. performed the readout SNR analysis via the construction of a rate equation model of the NV center, and utilized functional analysis to estimate the optimal time-varying laser waveform for spin readout \cite{liu2021pulsed}. However, these works highly rely on the modeling of the NV excited-state level structure and the transition rates between the levels, for which the corresponding parameters are difficult to be precisely determined and are susceptible to environmental influences in different applications. Therefore, these semi-theoretical optimization methods are generally not effective, leading to a hard-to-reach optimal performance in practice. Recently, Oshnik $et$ $al$. propose a two-step optimization strategy to improve the sensitivity of NV magnetometry by optimizing the squared-shape laser \cite{oshnik2022mag}.

Here, we propose a Hooke-Jeeves algorithm-assisted online optimization (OLO) method as a universal and efficient way to improve the performance of NV spin polarization and readout by customizing amplitude-modulated lasers. The scheme is inspired by the hybrid quantum-classical approach (HQCA) \cite{bauer2016HQCA,bravyi2016TradingComputationalResources,mcclean2016theoryHQCA} where the target spin acts as a provider of feedback information, and is controlled by a classical computer that runs an optimization algorithm in real-time. Since the scheme makes decisions directly based on feedback from target spin, it neither requires an accurate model of the target quantum system nor the specific optimization processes for the system. By applying this scheme to a NV-based sensor, we demonstrate an optimized photon time traces measurement and show a 44.1\% improvement in the SNR compared with the traditional scheme. As an application of this method, we improve the 
\begin{figure*}[ht!]
	\centering
	\includegraphics[width=\linewidth]{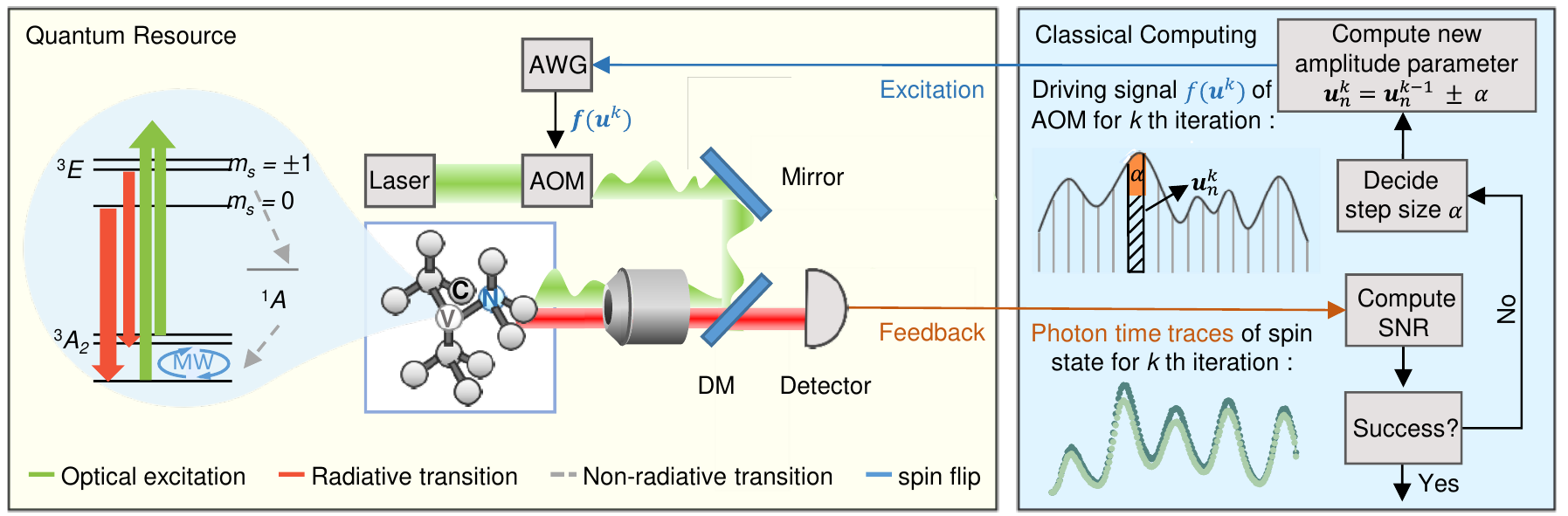}
	\caption{Illustration of algorithm-assisted OLO method for NV-based metrology. On the left is a schematic of the experimental setup, which mainly includes an AOM driven by the AWG and a 532 nm continuous-wave laser modulated by the AOM. The modulated laser is used for optical pumping of the electron spin in diamond. A detector records the fluorescences emitted by the spin system and feeds the photon time traces to the classical computer on the right. SNR is extracted from the feedback information, and subsequently utilized to generate a new set of laser parameters $\bm{u} = [u_1, u_2,..., u_n]$ by the algorithm. These parameters are then performed in new experiments for the next-round feedback information.}
	\label{fig:01}
\end{figure*}
contrast and mean deviation of the Rabi oscillation experiments by 46.0\% and 12.1\%, respectively. These experimental results demonstrate the feasibility of this scheme with state-of-the-art NV-based quantum sensing technologies and the scheme may also be applied to other quantum systems such as NV ensembles.

\section{ENERGY LEVELS AND OPTICAL PUMPING PROCESSES OF NV CENTER}
The NV center is a kind of point defect in diamonds consisting of a lattice vacancy and an adjacent nitrogen substitution \cite{suter2017magnetic,doherty2013nitrogen}. A simple model of the negatively charged NV center energy-levels is shown in Fig.\ref{fig:01} \cite{doherty2011negatively}. The transition between the ground triplet state $^{3}A_{2}$ and the excited triplet state $^{3}E$ enables optical addressability of NV centers with our homebuilt confocal microscopy system \cite{chen2020detecting}. The ground state NV center is excited with a 532 nm laser, while the excited state releases fluorescences by radiative relaxation to the ground state with the spin conserved \cite{fuchs2010Excited-stateSpinCoherence}. However, the population on $m_{s}=\pm1$ state has a higher probability than on $m_{s}=0$ to non-radiatively cross over from the triplet into the singlet state, which is called intersystem-crossing (ISC) \cite{goldman2015photon}. After being shelved for about 250 ns on the metastable state $^{1}E$, the spin preferentially returns to $m_{s}=0$ of the ground state \cite{Sellars2006dynamics}. The key here for spin-state optical readout is that, the NV center on $m_{s}=0$ state fluoresces more photons than on $m_{s}=\pm1$ state. During the readout process, the NV center is also installed to $m_{s}=0$ state after several optical cycles.

\begin{figure}[htbp]
	\centering
	\includegraphics[width=\linewidth]{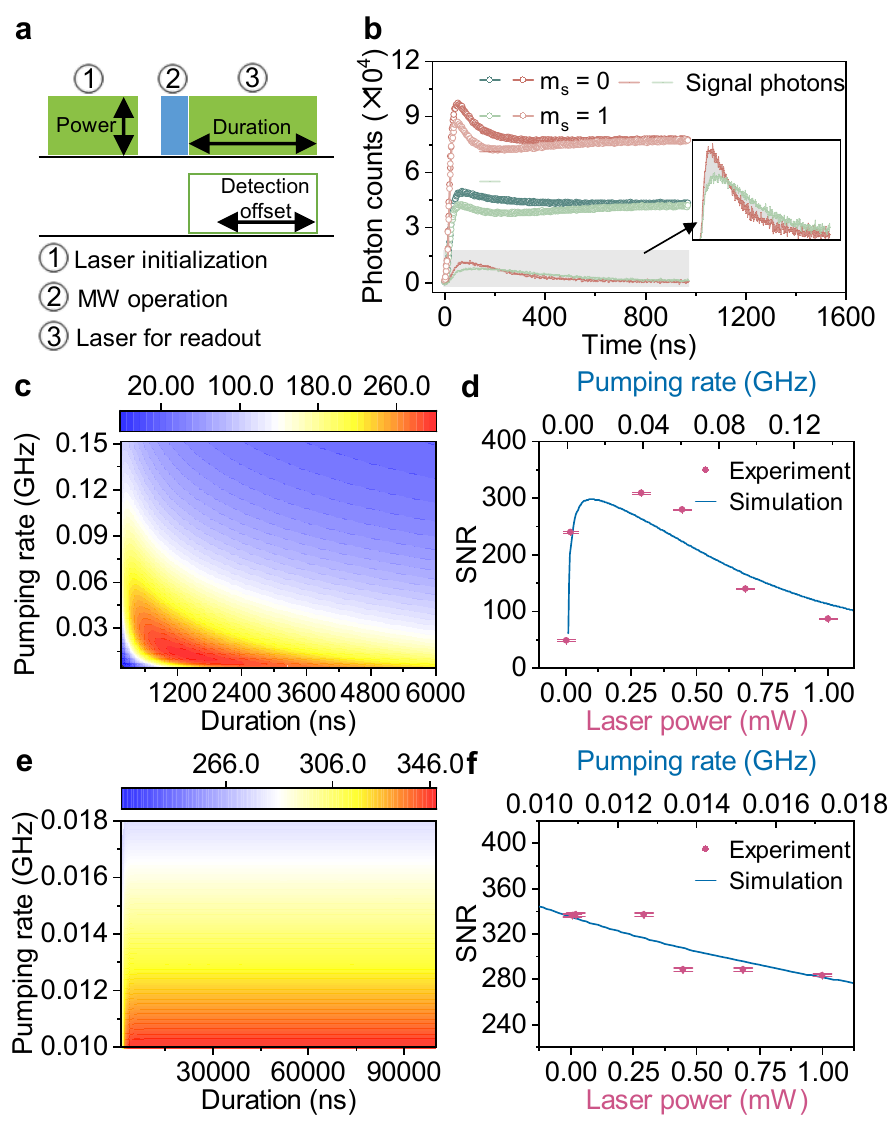}
	\caption{Analysis of optical pumping for the electron spin in diamond. (a) Pulse sequence of the electron-spin initialization and readout. The microwave operation only acts on the preparation of spin states other than $m_{s}=0$. (b) Photon time traces of spin states initialized at $m_{s}=0$ (dark green and dark red dotted lines) and $m_{s}=1$ (light green and light red dotted lines), measured with 10$^{8}$ repetitions. The red and green lines covered by the shadow are their differential signals, namely, signal photons. The inset plot shows the detailed comparison between signal photons at different detection offset and power of laser. (c) and (e) Theoretical SNR of the traditional scheme as a function of the pumping rate and duration for global optical pumping (including initialization and readout) and initialization, respectively. (d) and (f) SNR of the traditional scheme as a function of pumping rate in simulations or power in experiments under optimal duration conditions for global optical pumping and initialization, respectively.}
	\label{fig:02} 
\end{figure}

Conventionally, a squared-shape laser is used to initialize and read out the spin state (see Fig.\ref{fig:02}(a), where typical photon time traces of the $m_{s}=0$ and $m_{s}=1$ are shown in Fig.\ref{fig:02}(b). Usually, they are detected within an appropriate detection offset after turning on the readout laser to reduce the photon counts from the initialized NV center (electron-spin state readout noise), and are accumulated with multiple measurements to suppress the particle nature of photons (photon shot noise). The reliability of an optical spin-state readout for the NV-based sensors, designated as the SNR, is a common metric to evaluate sensing performance and compare different readout methods. The SNR for an optical measurement distinguishing the two possible spin projections $m_{s}=0$ and $m_{s}=\pm1$ of the NV center is defined as \cite{steiner2010fidelity}
\begin{equation}
{\rm SNR} = \frac{L_0 - L_1}{\sqrt{L_0 + L_1}}.
\label{eq:01}
\end{equation}
Here, $L_0 (L_1)$ is the total number of the collected photons in the detection offset when preparing the NV center in the spin projection $m_{s}=0$ ($m_{s}=\pm1$). The difference $L_0 - L_1$ represents the amount of signal photons, while the photon shot noise is the square root of the total number of collected photons, denoted by $\sqrt{L_0 + L_1}$ \cite{wolf2015readout,jg1946frequencydistribution}.

Adequate initialization, which is a prerequisite for high-quality subsequent manipulations, coupled with high readout SNR during the optical pumping guarantees an excellent sensing sensitivity. Given that the effect of initialization will eventually be reflected in the performance of readout, we consider merely the metric of readout SNR during optical pumping. Here, we analyze the effects of power and detection offset of the laser on optical pumping. On the one hand, it can be seen from Fig.\ref{fig:02}(b) that the photon time traces of $m_{s}=0$ and $m_{s}=1$ gradually coincide as the spin system is optically polarized, which leads to a gradual decrease in the number of signal photons. On the other hand, the increased laser power can lead to more photon counts and an issue of charge state instability due to ionization, but not necessarily more useful signal photons, because the readout noise (including photon scatter noise and electron spin state readout noise) is proportional to the laser power. The physical picture of the mechanism of our model can be explained as follows: Due to the ISC mechanism, the laser used for pumping will increase the readout noise when NV center is on the excited state or metastable state. Thus, an amplitude-modulated laser can reduce unnecessary laser illumination, meanwhile lowering the readout noise.

\section{FRAMEWORK}
To improve optical pumping for NV-based metrology by suppressing the readout noise and maximizing the usage of spin-related fluorescence, our new method is inspired by the HQCA to customize the laser waveform. The HQCA exploits the particular advantages of quantum processors and classical computers to jointly perform gradient-based optimizations, which has been demonstrated in many quantum information processing tasks, such as enhanced quantum metrology \cite{yang2021HQCA}, many-body state preparation \cite{li2017HQCAquantumoptimalcontrol,lu2017quantumcontrol12qubits} and quantum state tomogUraphy \cite{xin2020HQCATomography}. Specifically, the cost function and gradient computations are accomplished on the quantum processor, while the remaining is done on a classical computer. Unlike the normal HQCA applications, gradient calculations are not involved in our optimization process, therefore the task of the target spin system is to return the feedback information to the classical computer.

In experiment, we consider two schemes to customize laser waveform for initialization and readout, and use SNR as the evaluation criterion. Here, we only focus on two parameters of the laser, namely, power and duration. In the first scheme (traditional scheme), lasers used for initialization and readout are of constant power, $i.e.$ squared-shape laser. Considering the different purposes of initialization and readout, the second one uses different amplitude-modulated lasers for the two optical dynamic processes. The duration of the laser pulse is roughly determined by the first scheme. Then, the laser pulse is equally divided into multiple pieces, and the power of each piece is independently customized. 

Since the regulation of laser is achieved by controlling the driving signal of the AOM, we use the amplitude parameters of the driving signal instead of the power parameters. In addition, we stipulate that the laser is always turned on within the detection offset, that is, the duration is equal to the width of the detection offset mentioned above.

As depicted in Fig.\ref{fig:01}, the algorithm-assisted OLO method primarily consists of the quantum system to be optimized and a classical computer to compute the SNR from the feedback information and perform iterations. Specifically, on the one hand, the AWG receives the amplitude parameters $\bm{u} = [u_1, u_2,..., u_n]$ from the classical optimization algorithm and outputs the corresponding driving signal $f(\bm{u})$ to AOM, thereby realizing a modulated laser. On the other hand, the electron spin in the NV center is initialized and read out using the amplitude-modulated laser. The end of the overall update means the beginning of the next round of iterations. As the iterations progress, the step size $\alpha$ to update parameter gradually decreases, the SNR gradually converges to an optimal value, and the optimization is completed. It should be noted that each iteration cycle contains SNR queries and parameter updates with at least $n$ + 1 and up to 2$n$ times. The number of laser pieces $\bm n$ and experimental repetitions together largely determine the time overhead require for optimization. Photon time traces strongly associated with spin states are recorded by the detector and transmitted to the classical computer as a feedback for calculating the SNR. The algorithm determines new amplitude parameters based on the SNR. Through the multiple iterations of the closed-loop process described above, the SNR eventually converges to some optimal value, at which the optimal laser waveform is determined. 
\begin{figure}[htbp]
	\centering
	\includegraphics[width=\linewidth]{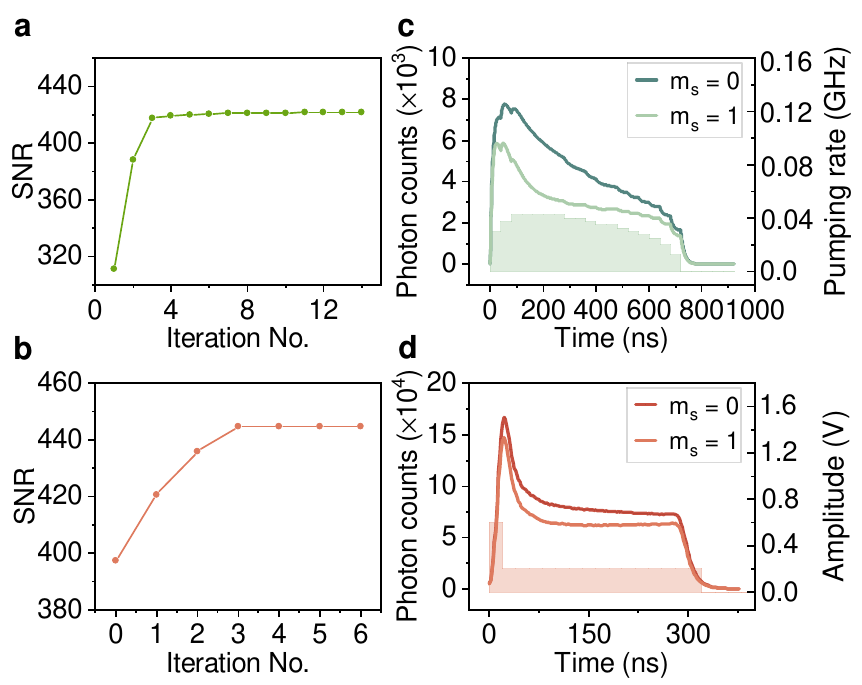}
	\caption{Simulation and experimental results of the OLO scheme for readout. (a) and (b) show the optimization process for simulations and experiments, respectively. The dotted lines represent the optimal SNR achieved at each iteration. (c) and (d) plot the final optimal laser waveform (shaded area) and photon time traces of the $m_{s}=0$ and $m_{s}=1$ for simulations and experiments, respectively.}
	\label{fig:03}
\end{figure}
In practice, we can use many different learning algorithms \cite{Jonathan2018Gradientascent,JieChen2021ML,Jogesh2020nn}, ranging from simple direct search algorithms \cite{lewis2000directsearchmethods} to more complex evolutionary algorithms \cite{eiben2015evolutionarycomputation}. Here, the classical Hooke-Jeeves direct search algorithm \cite{Hooke1961} is used to assist the OLO scheme with the optimization task for optical pumping. The entire online optimization process can be summarized as follows: In each iteration, the algorithm independently adjusts the parameter of each piece $u_i, i = 1,2,..,n$ sequentially and then an experiment is conducted to query the SNR. After the adjustment of all parameters, there is an overall update for the parameters, the direction of which is determined by the starting parameters of the round iteration pointing to the current optimal parameters. The details of the  Hooke-Jeeves algorithm are shown in the supplementary information.

\section{SIMULATION AND EXPERIMENTAL RESULTS}
To fully explore the effect of laser parameters on optical pumping, the method of parameter traversal is employed in the first scheme. We perform parameter traversal for the pumping rate and duration in simulations, power and duration in experiments to obtain the best laser waveform for SNR. Using the pumping rate instead of power in the simulation is reasonable because it is difficult to numerically correspond the pumping rate to the power, and they are positively correlated and proportional in a short range of steady-state. 
Fig. \ref{fig:02}(c) shows the theoretical relationship between laser parameters and SNR in the traditional scheme. For better visualization, the two-dimensional plot is projected onto the pumping rate to obtain the optimal SNR under different pumping rates, as shown in the blue curve in Fig.\ref{fig:02}(d). The optimal SNR with traditional scheme in simulation and experiment is 298.2 and 308.6, respectively. From the blue theoretical curve and the pink experimental dots in Fig.\ref{fig:02}(d), both theoretical and experimental results are consistent in showing that the SNR acquires an optimal value as pumping rate in simulations or power in experiments increases, which is due to the increasing proportion of the readout noise.

Based on the optimal laser parameters given by the traditional scheme, initialization and readout for NV-based metrology are optimized with the OLO scheme. In fact, the amplitude-modulated laser with multi-piece is not indispensable for initialization, which focuses on polarization rather than noise suppression. So we pay more attention to the optimization of spin readout, setting the number of laser pieces $n=1$ for initialization and $n$ = 20 for readout in experiments. Fig. \ref{fig:02}(e) shows that the theoretical and experimental results between the laser parameters used for initialization and the SNR in the OLO scheme. Similarly, the two-dimensional plot is projected onto the pumping rate to obtain the optimal SNR under different pumping rates, as shown in the blue line in Fig.\ref{fig:02}(f). It can be seen that the experimental results (pink dots) are roughly identical to the simulations. 

With the optimal parameters used for initialization, a random starting parameters (duration, pumping rate or amplitude) for readout optimization in simulation and experiment is chosen to be (920 ns, 0.01 GHz) and (390 ns, 0.2 V), respectively. Here, only one value of pumping rate or amplitude is given because the initial value is the same for each pieces. Note that the duration does not change during optimization. It is used in optical pumping with the starting pumping rate or amplitude (denoted as $\bm{u}^{0}$) to query the initial SNR. Then, by feeding both the starting pumping rate or amplitude and SNR into the Hooke-Jeeves algorithm, we can systematically produce new pumping rate or amplitude parameters. The pumping rates in simulation and the amplitude parameters in experiment vary in the range of [0.0, 0.5] (GHz) and [0, 1] (V). We emphasize that the entire procedure was fully automated, such that this procedure can proceed ad infinitum without intervention till stopping conditions are met. The dotted lines in Fig.\ref{fig:03}(a) and \ref{fig:03}(b) show the theoretical and experimental optimization process by the algorithm-assisted OLO method, respectively. The searched optimal parameters (see Fig.\ref{fig:03}(c) and \ref{fig:03}(d)) induce a final SNR of 421.9 in simulations and 444.7 in experiments, which are 41.8\% and 44.1\% higher than using the constant waveform (the traditional scheme), respectively.

\section{APPLICATION}
\begin{figure}[htbp]
	\centering
	\includegraphics[width=\linewidth]{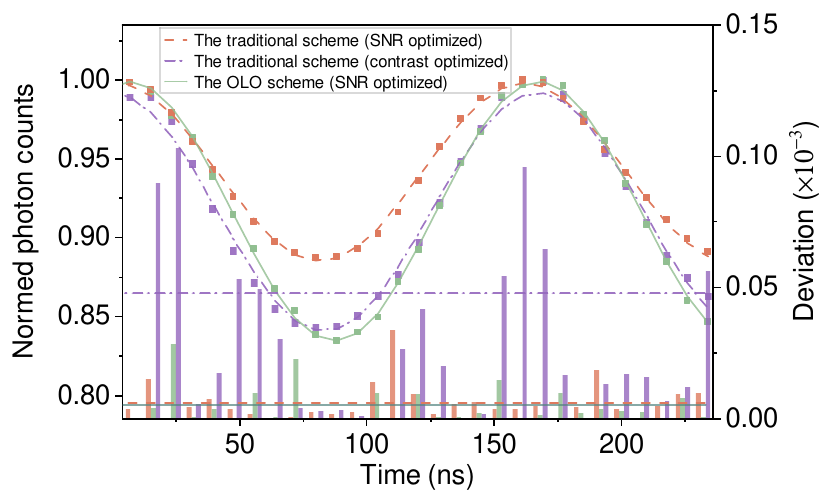}
	\caption{Rabi oscillation measurements (left axis) and their deviation (right axis) with different optimizations. The fitted contrast of the OLO scheme (green, solid line), the traditional scheme optimized by SNR (orange, dashed line), and the traditional scheme optimized by constrast (purple, dash-dot line) are 16.5\%, 11.3\%, and 15.2\%, respectively. Columns are deviations of Rabi measurements, defined by differences between the data points and the best fitted lines. The means of deviations on each Rabi oscillation, are 0.00538 (OLO scheme by SNR), 0.00612 (traditional scheme by SNR), and 0.04794 (traditional scheme by contrast), respectively. Both contrast and deviation are improved by our OLO scheme.}
	\label{fig:04}
\end{figure}
Here, we apply the OLO method to the Rabi oscillation of an electron spin driven under resonant microwave (MW) magnetic field, which is a scheme to measure both the strength and orientation of the MW magnetic field \cite{golter2014Rabi,robledo2010Rabi}. In Rabi oscillation measurement, the spin state can be oscillated from the $m_{s}=0$ ground state into the $m_{s}=1$ state under the action of MW. Fig. \ref{fig:04} shows the standard Rabi oscillation curves and the fluorescence data are normalized in units of $m_{s}=0$ state fluorescences. The measurement of each point on the curve involves the optical pumping processes related to initialization and readout. In Rabi experiments, a common metric used for determining laser parameters is contrast, which is defind as \cite{hopper2018readout} 
\begin{equation}
{\rm C} = \frac{L_0 - L_1}{L_0}.
\label{eq:02}
\end{equation}
However, changing the laser parameters in the traditional scheme cannot optimize contrast while ensuring the reliability of readout.
As shown in Fig.\ref{fig:04}, the contrast of purple Rabi curve (contrast optimized) reaches 15.2\%, while that of orange Rabi curve (SNR optimized) drops to 11.3\%, but the data points in orange Rabi curve are much better aligned on a sinusoidal curve. For better visualization, Fig.\ref{fig:04} displays the differences between the points and the best fitted lines. The data deviation can originate from readout noise, MW pulse errors, spin-mixing, and so on. The general reduction in the data deviation indicates a more accurate spin state readout.
With the algorithm-assisted OLO scheme, the contrast of Rabi experiments is improved by 46.0\% and 8.6\% compared to the traditional method that aims to improving SNR and contrast, respectively. Meanwhile, the mean deviation has been witnessed a 12.1\% and 88.8\% reduction compared to the traditional method.

\section{EXPERIMENT SETUP}
To demonstrate, we use a home-built optically detected magnetic resonance (ODMR) system to address and manipulate the single NV center in a $^{12}$C-purified single-crystal diamond, where the density of NV center is on the order of $10^{12}$ ${\rm cm^{-3}.}$ The Single NV center used for the experiments is located at depth of 7 ${\rm \mu}$m and its dephasing time $T_2^*$ is about 10 $\mu$s. As shown in Fig. \ref{fig:01}, the 532-nm green laser beam is modulated by a 200 MHz acoustic-optic modulator (AOM) before passing through a 100X oil objective and focusing on the NV center. Different from the normal driver, the AOM is driven by a 2.6 GS/s arbitrary wave generator (Tektronix AWG610) with an output bandwidth over 800 MHz, which is capable of generating laser pulses as short as 4 ns. The fluorescence is emitted through the objective and collected by an avalanche photodiode (APD). The output signal of APD is detected by a time tagger for a time-resolved measurement. A static magnetic field around 50 mT is applied along the NV axis to split the sub-levels and the associated nitrogen nuclear spin is polarized at the excited-state level anticrossing as well. The microwave source (Rohde$\&$Schwartz SMIQ03B) is used to manipulate the electron-spin state for general purpose measurements. An impedance-matched copper slot line with an $\Omega$-type ring is used to radiate the microwave fields. All the electronic and microwave devices are synchronized by a 10 MHz atomic clock source.

\section{DISCUSSION AND CONCLUSION}
In summary, by combining the classical algorithm, an online optimization method (algorithm-assisted OLO method) of optical pumping is proposed to enhance the sensitivity of quantum metrology. To improve the polarization and suppress readout noise during optical pumping, a single laser pulse is divided into multiple pieces, and the power of each piece is independently determined by the algorithm-assisted OLO method. This method requires additional hardware that supports waveform editing to realize an amplitude-modulated laser. In addition to the arbitrary waveform generator (AWG) used in our experiments, the low-cost direct digital synthesizer chip (DDS) can also meet the needs of the OLO method. Compared to the traditional optical pumping scheme, our new scheme improves the SNR of NV metrology by 44.1\%. In addition, we achieve 46.0\% increase in contrast and 12.1\% reduction in mean deviation in the Rabi measurement with the OLO method. High readout contrast and short total overhead time contribute to realizing high sensitivity. Considering the total overhead in experiments, our method requires an extra time cost of approximately a few microsecond to perform the laser pulses compared with the traditional scheme. In most NV-based applications, such as magnetic signals sensing and quantum register, the interrogation time ranges from hundreds of microseconds to a few milliseconds \cite{pham2012metrology,lukin2007register,Pham2011imaging}. The increase in initialization and readout overhead (\textless1\%) has little effect on sensitivity compared with increased readout contrast due to the enhanced performance in initialization and readout.

In this work, we pay more attention to the improvement of the readout, so only simple optimization is made for initialization. A previous study \cite{song2020pulse} shows that using repeatedly applied short laser pulses to initialize electron spin will further improve the electron spin readout efficiency. To conclude, our new method is beneficial to optimizing the NV metrology and other high-precision measurements. Thanks to the closed-loop optimization process in the method, we do not need to know the specific model of the system to be optimized in advance, so it can be easily extended to other quantum sensing platforms such as NV ensembles.\\

\begin{acknowledgments}
This work was supported by the National Key R\&D Program of China (Grants No. 2018YFA0306600 and 2019YFA0308100), the National Natural Science Foundation of China (Grant No. 92265114, 92265204 and 11875159), and the Research Initiation Project (No. K2022MB0PI02) of Zhejiang Lab.
\end{acknowledgments}

\nocite{*}

\bibliography{ref_main}

\end{document}